\documentclass[12pt]{article}

\newcommand{\be}{\begin{equation}}
\newcommand{\ee}{\end{equation}}
\newcommand{\prt}{\partial}
\newcommand{\om}{\omega}
\newcommand{\sgm}{\sigma}
\newcommand{\al}{\alpha}
\newcommand{\gm}{\gamma}
\newcommand{\ra}{\rightarrow}
\newcommand{\ep}{\varepsilon}

\begin{document}

\begin{center}
{\Large{\bf Collective Phenonema in the Interaction of Radiation with 
Matter}} \\ [5mm]

{\large V.I. Yukalov} \\ [3mm]

{\it Insituto de Fisica de S\~ao Carlos, Universidade de S\~ao Paulo \\
Caixa Postal 369, S\~ao Carlos, S\~ao Paulo 13560--970, Brazil}
\end{center}

\begin{center}

\vskip 5mm

{\large{\bf Abstract}}

\end{center}

The aim of this communication is to present in a concentrated form the
main ideas of a method, developed by the author, for treating strongly 
nonequilibrium collective phenomena typical of the interaction of radiation 
with matter, as well as to give a survey of several applications of the 
method. The latter is called the Scale Separation Approach since its 
basic techniques rely on the possibility of separating different 
space--time scales in nonequilibrium statistical systems. This approach 
is rather general and can be applied to diverse physical problems, 
several of which are discussed here. These problems are: Superradiance of 
nuclear spins, filamentation in resonant media, semiconfinement of 
neutral atoms, negative electric current, and collective liberation of
light.

\vskip 2cm

\section{Introduction}

\begin{sloppypar}

Strongly nonequilibrium processes that occur in statistical systems
and involve their interaction with radiation are usually described by 
complicated nonlinear differential and integro--differential equations
[1--3]. For treating these difficult problems, a novel approach has recently
been developed [4--7] called the {\it scale separation approach} since its
main idea is to formulate the evolution equations in such a form where it
could be possible to separate several characteristic space--time scales.
In many cases, different scales appear rather naturally being directly
related to the physical properties of the considered system.

The scale separation approach has been employed for solving several 
interesting physical problems related to strongly nonequilibrium 
processes occurring under the interaction of radiation with matter.
As an illustration, the following phenomena are selected for this report:
{\it Superradiance of Nuclear Spins}, {\it Filamentation in Resonant 
Media}, {\it Semiconfinement of Neutral Atoms}, {\it Negative Electric
Current}, and {\it Collective Liberation of Light}.

Since the scale separation approach makes the mathematical foundation for 
the following applications, its general scheme is described in Section 2.
In Sections 3 to 7 concrete physical effects are briefly reviewed 
and the most important results are summarized.

\section{Scale Separation Approach}

Because of the pivotal role of this approach for treating different 
physical problems, its general scheme will be presented here in an 
explicit way [4--7]. It is possible to separate the following main steps,
or parts, of the approach.

\subsection{Stochastic quantization of short--range correlations}

When considering nonequilibrium processes in statistical systems, one needs 
to write evolution equations for some averages $<A_i>$ of operators 
$A_i(t)$ where $t$ is time and $i=1,2,\ldots, N$ enumerates particles 
composing the considered system. For simplicity, a discrete index $i$ is 
used, although everywhere below one could mean an operator $A(\vec 
r_i,t)$ depending on a continuous space variable $\vec r_i$.

There is the well known problem in statistical mechanics consisting in the 
fact that writing an evolution equation for $<A_i>$ one does not get a 
closed system of equations but a hierarchical chain of equations 
connecting correlation functions of higher orders. Thus, an equation for 
$<A_i>$ contains the terms as $\sum_j<A_iB_j>$ with double correlators 
$<A_iB_j>$, and the evolution equations for the latter involve the terms 
with tripple correlators, and so on. The simplest way for making the 
system of equations closed is the mean--field type decoupling $<A_iB_j>\ra
<A_i><B_j>$. When considering radiation processes, this decoupling is 
called the semiclassical approximation. Then the term $\sum_j<A_iB_j>$ 
reduces to $<A_i>\sum_j<B_j>$, so that one can say that $<A_i>$ is 
subject to the action of the mean field $\sum_j<B_j>$. The semiclassical 
approximation describes well coherent processes, when long--range 
correlations between atoms govern the evolution of the system, while 
short--range correlations, due to quantum fluctuations, are not 
important. However, the latter may become of great importance for some 
periods of time, for example, at the beginning of a nonequilibrium 
process when long--time correlations have had yet no time to develop. 
Then neglecting short--range correlations can lead to principally wrong 
results.

To include the influence of short--range correlations, the semiclassical 
approximation can be modified as follows:
\be
\label{1}
\sum_j <A_iB_j> \; = \; <A_i>\left ( \sum_j <B_j> \; + \xi \right )\; ,
\ee
where $\xi$ is a random variable describing local short--range 
correlations. It is natural to treat $\xi$ as a Gaussian stochastic 
variable with the stochastic averages
\be
\label{2}
\ll\xi\gg\; = 0 \; , \qquad \ll |\xi|^2\gg\; = \; \sum_j |<B_j>|^2 \; ,
\ee
where the second moment is defined so that to take into account 
incoherent local fluctuations. Since short--range correlations are often 
due to quantum fluctuations, the manner of taking them into account by 
introducing a stochastic variable $\xi$ can be called the stochastic 
quantization. Then the decoupling (1) may be termed the {\it stochastic
semiclassical approximation}. This kind of approximation has been used 
for taking into account quantum spontaneous emission of atoms in the 
problem of atomic superradiance.

\subsection{Separation of solutions onto fast and slow}

The usage of the stochastic semiclassical approximation makes it possible 
to write down a closed set of stochastic differential equations. The next 
step is to find such a change of variables which results in the 
possibility of separating the functional variables onto fast and slow, so 
that one comes to the set of equations having the form
\be
\label{3}
\frac{du}{dt} = f(\ep,u,s,\xi,t) \; , \qquad
\frac{ds}{dt} = \ep\; g(\ep,u,s,\xi,t) \; ,
\ee
where $\ep\ll 1$ is a small parameter, such that
\be
\label{4}
\lim_{\ep\ra 0}\; f\neq 0\; , \qquad \lim_{\ep\ra 0}\; \ep\; g = 0\; .
\ee
As is evident, dealing with only two functions, $u$ and $s$, and one 
small parameter $\ep$ is done just for simplicity. All procedure is 
straightforwardly applicable to the case of many functions and several 
small parameters.

From Eqs. (3) and (4) it follows that
\be
\label{5}
\lim_{\ep\ra 0}\; \frac{du}{dt}\neq 0 \; , \qquad
\lim_{\ep\ra 0}\; \frac{ds}{dt} = 0 \; ,
\ee
which permits one to classify the solution $u$ as fast, compared to the 
slow solution $s$. In turn, the slow solution $s$ is a {\it 
quasi--invariant} with respect to the fast solution $u$.

The above classification of solutions onto fast and slow concerns time 
variations. In the case of partial differential equations, one has, in 
addition to time, a space variable $\vec r$. Then the notion of fast or 
slow functions can be generalized as follows [8,9]. Let $\vec r\in {\bf 
V}$, with ${\rm mes}{\bf V}\equiv V$, and $t\in [0,T]$, where $T$ can be 
infinite. Assume that
\be
\label{6}
\lim_{\ep\ra 0}\; \ll \frac{1}{V}\int_{\bf V} \frac{\prt u}{\prt t} \;
d\vec r \gg \; \neq 0 \; , \qquad
\lim_{\ep\ra 0}\; \ll \frac{1}{T}\int_0^T \vec\nabla u \;
dt \gg \; \neq 0 \; ,
\ee
while
\be
\label{7}
\lim_{\ep\ra 0}\; \ll \frac{1}{V}\int_{\bf V} \frac{\prt s}{\prt t} \;
d\vec r \gg \; = 0 \; , \qquad
\lim_{\ep\ra 0}\; \ll \frac{1}{T}\int_0^T \vec\nabla s\;
dt \gg \; = 0 \; .
\ee
Then the solution $u$ is called {\it fast on average}, with respect to 
both space and time, as compared to $s$ that is {\it slow on average}. In 
such a case $s$ is again a quasi--invariant as compared to $u$. In 
general, it may, of course, happen that one solution is fast with respect 
to time but slow in space, or vice versa, when compared to another 
function. The notion of quasi--invariants with respect to time is known 
in the Hamiltonian mechanics where they are also called adiabatic 
invariants. Here this notion is generalized to the case of both space and 
time variables [8,9].

\subsection{Averaging method for multifrequency systems}

After classifying in Eqs. (4) the function $u$ as fast and $s$ as slow, 
one can resort to the Krylov--Bogolubov averaging technique [10] extended 
to the case of multifrequency systems. This is done as follows.

Since the slow variable $s$ is a quasi--invariant for the fast variable
$u$, one considers the equation for the fast function $u$, with the slow 
one kept fixed,
\be
\label{8}
\frac{\prt X}{\prt t} = f(\ep,X,z,\xi,t) \; .
\ee
Here $z$ is treated as a fixed parameter. The solution to Eq. (8), that is
\be
\label{9}
X = X(\ep,z,\xi,t) \; , \qquad z =const \; ,
\ee
has to be substituted into the right--hand side of the equation for the 
slow function, and for this right--hand side one defines the average
\be
\label{10}
\overline g(\ep,z) \equiv \; \ll \frac{1}{\tau} \int_0^\tau 
g(\ep,X(\ep,z,\xi,t),z,\xi,t) \; dt \gg \; ,
\ee
in which $\tau$ is the characteristic oscillation time of the fast 
function. In many cases, it is possible to take $\tau\ra\infty$, 
especially when the period of fast oscillations is not well defined [2]. 
Then one comes to the equation
\be
\label{11}
\frac{dz}{dt} = \ep\; \overline g(\ep,z)
\ee
defining a solution
\be
\label{12}
z=z(\ep,t) \; .
\ee
Substituting the latter into $X$, one gets
\be
\label{13}
y(\ep,\xi,t) = X(\ep,z(\ep,t),\xi,t) \; .
\ee

The pair of solutions (9) are called the {\it generating solutions} since 
these are the first crude approximations one starts with. More elaborate 
solutions are given by Eqs. (12) and (13) which are termed {\it guiding 
centers}.

Notice two points that difference the case considered from the usual 
averaging techniques. The first point is that in Eq. (8) the small 
parameter $\ep$ is not set zero. And the second difference is in the 
occurrence of the stochastic average in Eq. (10). Leaving $\ep$ in Eq. 
(8) makes it possible to correctly take into account attenuation effects, 
as will be shown in applications.

\subsection{Generalized expansion about guiding centers}

Higher--order corrections to solutions may be obtained by presenting the 
latter as asymptotic expansions about the guiding centers (12) and (13). 
To this end, $k$--order approximations are written as
$$
u_k = y(\ep,\xi,t) + \sum_{n=1}^k y_n(\ep,\xi,t)\; \ep^n\; ,
$$
\be
\label{14}
s_k = z(\ep,t) + \sum_{n=1}^k z_n(\ep,\xi,t)\; \ep^n\; .
\ee
Such series are called {\it generalized asymptotic expansions} [11] since 
the expansion coefficients depend themselves on the parameter $\ep$. The 
right--hand sides of Eqs. (3) are to be expanded similarly to Eq. (14) 
yielding
\be
\label{15}
f(\ep,u_k,s_k,\xi,t) \simeq f(\ep,y,z,\xi,t) +
\sum_{n=1}^k f_n(\ep,\xi,t) \; \ep^n
\ee
and an equivalent expansion for $g$. These expansions are to be 
substituted into Eqs. (3) with equating the like terms with respect to 
the powers of $\ep$. In the first order, this gives
\be
\label{16}
\frac{dy_1}{dt} = f_1(\ep,\xi,t) - \overline g(\ep,z) X_1(\ep,\xi,t)\; ,
\qquad
\frac{dz_1}{dt} = g(\ep,y,z,\xi,t) - \overline g(\ep,z)\; ,
\ee
where
$$
X_1(\ep,\xi,t) \equiv \frac{\prt}{\prt z}\; X(\ep,z,\xi,t) \; , \qquad 
z=z(\ep,t) \; .
$$
For the approximations of order $n\geq 2$, one gets
\be
\label{17}
\frac{dy_n}{dt}  = f_n(\ep,\xi,t) \; , \qquad
\frac{dz_n}{dt}  = g_n(\ep,\xi,t) \; .
\ee
The functions $f_n$ and $g_n$ depend on $y_1,y_2,\ldots,y_n$ and on 
$z_1,z_2,\ldots,z_n$ (see for details [6,7]). But it is important that the 
dependence on $y_n$ and $z_n$ is linear. Therefore all equations (16) and
(17) are linear and can be easily integrated. Thus, the approximants (14) 
are defined. Each $k$--order approximation can also be improved by 
invoking the self--similar summation of asymptotic series [12--18].

\subsection{Selection of scales for space structures}

The solutions of differential or integro--differential equations in 
partial derivatives are often nonuniform in space exhibiting the 
formation of different spatial structures. Also, it often happens that a 
given set of equations possesses several solutions corresponding to 
different spatial patterns or to different scales of such patterns [3]. 
When one has a set of solutions describing different 
possible patterns, the question arises which of these solutions, and 
respectively patterns, to prefer? This problem of pattern selection is a 
general and very important problem constantly arising in considering 
spatial structures. In some cases this problem can be solved as follows.

Assume that the obtained solutions describe spatial structures that can 
be parametrized by a multiparameter $b$, so that the $k$--order 
approximations $u_k(b,t)$ and $s_k(b,t)$ include the dependence on $b$ 
whose value is however yet undefined. To define $b$, and respectively the 
related pattern, one may proceed in the spirit of the self--similar 
approximation theory [12--14], by treating $b$ as a control function, or a 
set of control functions if $b$ is a multiparameter. According to the theory 
[12--14], control functions are to be defined from fixed--point conditions 
for an approximation cascade, which is to be constructed for an observed 
quantity. For the latter, one may take the energy which is a functional 
$E[u,s]$ of the solutions. In experiments, one usually measures an average 
energy whose $k$--order approximation writes
\be
\label{18}
E_k(b) \equiv \; \ll \frac{1}{\tau}\; \int_0^\tau E[u_k(b,t),s_k(b,t)]\;
dt \gg \; ,
\ee
where $\tau$ is a period of fast oscillations. For the sequence of 
approximations, $\{ E_k(b)\}$, it is possible to construct an 
approximation cascade [12--14] and to show that its fixed point can be given 
by the condition
\be
\label{19}
\frac{\prt}{\prt b} E_k(b) = 0  \; ,
\ee
from which one gets the control function $b=b_k$ defining the corresponding 
pattern. According to optimal control theory, control functions are defined 
so that to minimize a cost functional. In this case, it is natural to take 
for the latter the average energy (18). Therefore, if the fixed--point 
equation (19) has several solutions, one may select of them that one 
which minimizes the cost functional (18),
\be
\label{20}
E_k(b_k) = {\rm abs}\min_b E_k(b) \; .
\ee
Equations (19) and (20) have a simple physical interpretation as the 
minimum conditions for the average energy (18). However, one should keep 
in mind that there is no in general such a principle of minimal energy 
for nonequilibrium systems [3]. Therefore the usage of the ideas from 
the self--similar approximation theory [12--14] provides a justification for 
employing conditions (19) and (20) for nonequilibrium processes.

In the following sections a brief survey is given of several physical 
examples the scale separation approach has been applied to, and the main 
results are formulated.

\section{Superradiance of Nuclear Spins}

A system of neutral spins in an external magnetic field , prepared in a 
strongly nonequilibrium state and coupled with a resonance electric 
circuit, displays rather nontrivial relaxation behaviour somewhat similar 
to that of an inverted system of atoms. This is why the optical 
terminology, such as superradiance, has been used for describing 
collective relaxation processes in nonequilibrium nuclear magnets
[5,6,19,20].

For a system of nuclear spins interacting through dipole forces the 
evolution equations can be derived [5,6] for the averages
\be
\label{21}
u \equiv \frac{1}{N} \sum_{i=1}^N < S_i^-> \; , \qquad 
s\equiv \frac{1}{N} \sum_{i=1}^N < S_i^z> \; ,
\ee
in which $N$ is the number of spins, angle brackets mean statistical 
averaging, $S_i^-$ is a lowering spin operator, and $S_i^z$ is the 
$z$--component of a spin operator. Following the ideology of the scale 
separation approach, local fluctuating fields are presented by stochastic 
variables $\xi_0$ and $\xi$. In this way, one comes to the evolution 
equations for the transverse spin variable
\be
\label{22}
\frac{du}{dt} = i (\om_0 - \xi_0 + i\gm_2) u - i (\gm_3 h + \xi) s
\ee
and the longitudinal average spin
\be
\label{23}
\frac{ds}{dt} = \frac{i}{2} (\gm_3h  + \xi) u^* - \frac{i}{2} 
(\gm_3 h + \xi^*) u -\gm_1 ( s-\zeta)\; .
\ee
It is also convenient to consider the equation
\be
\label{24}
\frac{d}{dt}|u|^2 = -2\gm_2 |u|^2  - i (\gm_3 h + \xi) su^* +
i(\gm_3 h + \xi^*) su \; .
\ee
In equations (22)--(24) dimensionless units are used for the resonator 
magnetic field $h$ satisfying the Kirchhoff equation
\be
\label{25}
\frac{dh}{dt} + 2\gm_2 h + \om^2\int_0^t h(t')\; dt' =
-2\al_0\;\frac{d}{dt}\; ( u^* + u) +\gm_3f \; .
\ee
Here $\om_0$ is the Zeeman frequency of spins in an external uniform 
magnetic field, $\om$ is the resonator natural frequency, $\gm_1$ and 
$\gm_2$ are the spin--lattice and spin--spin relaxation parameters, 
respectively, $\gm_3$ is the resonator ringing width, $\zeta$ is a 
stationary spin polarization, $\al_0$ is the coupling between spins and 
the resonator, and $f$ is an electromotive force. The random local fields 
are defined as Gaussian stochastic variables with the stochastic averages
\be
\label{26}
\ll \xi_0^2\gg \; = \; \ll |\xi|^2\gg\; = \gm_2^* \; ,
\ee
where $\gm_2^*$ is the inhomogeneous dipole broadening.

There are the following small parameters in the system:
$$
\frac{\gm_1}{\om_0} \ll 1\; , \qquad
\frac{\gm_2}{\om_0} \ll 1\; , \qquad
\frac{\gm_2^*}{\om_0} \ll 1\; , \qquad
\frac{\gm_3}{\om} \ll 1\; , 
$$
\be
\label{27}
\frac{\Delta}{\om_0} \ll 1\; , \qquad (\Delta\equiv \om -\om_0) \; .
\ee
This makes it admissible to classify the functions $u$ and $h$ as fast, 
while $s$ and $|u|^2$ as slow, and to apply the method of Section 1. The 
behaviour of solutions to Eqs. (22)--(25) depends on initial conditions 
for $u(0)$, and $s(0)$, on the existence of an electromotive driving 
force $f(t)$, on the pumping related to the parameter $\zeta$, and on the 
value of the effective coupling parameter
\be
\label{28}
g =\pi^2\eta\; \frac{\rho\;\mu_n^2\;\om_0}{\hbar\;\gm_2\;\om} \; ,
\ee
in which $\eta$ is a filling factor; $\rho$, spin density; and $\mu_n$ is 
a nuclear magnetic moment.

The first interesting result is that the electromotive force does not 
influence much macroscopic samples [5,6] since the corresponding 
correlation time is proportional to $N$, that is, the effective, 
interaction strength of an electromotive force with the spin system is 
proportional to $N^{-1}$. This shows, in particular, that the role of the 
thermal Nyquist noise for starting the relaxation process is negligible. 
The main cause triggering the motion of spins leading to coherent 
self--organization is the presence of {\it nonsecular dipole interactions}
[5,6,19]. The latter result gives an answer to the problem, posed
by Bloembergen and Pound [21]: What is the origin of self--organized 
coherent relaxation in spin systems?

All possible regimes of nonlinear spin dynamics have been analysed 
[5,6,19,20]. When the nonresonant external pumping is absent, that is 
$\zeta>0$, there are seven qualitatively different transient relaxation 
regimes: {\it free induction, collective induction, free relaxation, 
collective relaxation, weak superradiance, pure superradiance}, and 
{\it triggered superradiance} [6]. In the presence of pumping, realized 
e.g. by means of dynamical nuclear polarization directing nuclear spins 
against an external constant magnetic field, one has $\zeta\leq 0$. Then 
three dynamical regimes can be observed, depending on the value of $\zeta$ 
with respect to the pumping thresholds
\be
\label{29}
\zeta_1 = -\frac{1}{g} \; , \qquad
\zeta_2 = -\frac{1}{g} \left ( 1 +\frac{\gm_1^*}{2\gm_2}\right ) \; ,
\ee
where $\gm_1^*$ is an effective pumping rate.

Two stationary points can exist for the slow solutions $s$ and $w$, where
$$
w\equiv |u|^2 - 2 \left ( \frac{\gm_2^*}{\om_0}\right )^2 s^2 \; .
$$ These fixed points are
$$
s_1^* =\zeta \; , \qquad w_1^* = 0 \; ,
$$
\be
\label{30}
s_2^* = - \frac{1}{g} \; , \qquad w_2^* = - 
\frac{\gm_1^*(1+g\zeta)}{g^2\gm_2}\; .
\ee
When $\zeta_1<\zeta\leq 0$, then the first fixed point is a stable node 
and the second one is a saddle point. For $\zeta=\zeta_1$, both 
stationary points merge together, being neutrally stable. After the 
bifurcation at the value $\zeta=\zeta_1$, in the region 
$\zeta_2\leq\zeta <\zeta_1$, the first fixed point looses its stability 
becoming a saddle while the second fixed point becomes a stable node. 
Finally, when $\zeta<\zeta_2$, the second fixed point transforms to a 
stable focus, and the first one is, as earlier, a saddle point.

In this way, there are three qualitatively different lasting relaxation 
regimes induced by the pumping. The first one is a monotonic relaxation 
to the first stationary solution with practically no coherence,
$w_1^*=0$. The second regime is a monotonic relaxation to the 
second stationary solution with a nonzero coherence, $w_2^*\neq 0$. And 
the third regime is that of pulsing relaxation to the coherent stationary 
point. Note that the pumping rate $\gm_1^*$ can be larger than $\gm_2$, 
so that $w_2^*$ can reach the order of unity. The three lasting 
relaxation regimes occurring in the presence of pumping can be called, 
respectively, {\it incoherent monotone attenuation, coherent monotone 
relaxation}, and {\it coherent pulsing relaxation}.

\section{Filamentation in Resonant Media}

In optical resonant media there appear space structures when the 
radiation wavelength is much less than the characteristic sizes of the 
laser system [22]. There are two principally different types of spatial
structures in laser media, one corresponding to low Fresnel numbers
[23--28] and another type corresponding to high Fresnel numbers [29--35],
with a transition occurring around $F\sim 10$. Similar effects are observed
in photorefractive media [36--38]. Such structures are described by nonlinear
differential equations in partial derivatives. The general problem in dealing
with these equations is the nonuniqueness of their solutions each of which
corresponds to a particular spatial structure [3]. The related problem 
of pattern selection can be treated by the method of subsection 2.5. Here 
this is illustrated by the theory of filamentation in optical resonant 
media [39--42] at high Fresnel numbers.

The Hamiltonian for a system of resonant atoms can be written as
$$
\hat H = \frac{1}{2}\sum_i \om_0 ( 1 +\sgm_i^2) - 
\frac{1}{2} \sum_i \left ( \vec d^*\cdot\vec E^+_{0i}\sgm_i^-
+ \sgm_i^+\vec d\cdot\vec E^-_{0i}\right ) -
$$
\be
\label{31}
-\frac{1}{2}\sum_{i\neq j} \left ( \vec d^*\cdot\vec E^+_{ij}\sgm_i^- +
\sgm_i^+\vec d\cdot\vec E^-_{ij} \right ) \; ,
\ee
where the standard notation is used for the Pauli matrices $\sgm_i^\pm$ and
$\sgm_i^z$; $\vec d$ is a transition dipole for a transition with the 
frequency $\om_0$; the electric fields
$$
\vec E^-_{0i} =\vec E_0 e^{i(kz_i -\om t)}  \qquad 
\left ( k\equiv \frac{\om}{c}\right ) \; ,
$$
\be
\label{32}
\vec E^-_{ij} = \frac{k_0^2}{r_{ij}}\; \vec n_{ij}\times \left ( \vec d\times
\vec n_{ij}\right ) e^{ik_0r_{ij}}\sgm^-_j \qquad
\left ( k_0 \equiv \frac{\om_0}{c}\right )
\ee
correspond to the laser mode and to the reradiated field, 
respectively, and
$$
r_{ij} \equiv |\vec r_{ij}| \; , \qquad 
\vec n_{ij} \equiv \frac{\vec r_{ij}}{r_{ij}} \; , \qquad 
\vec r_{ij} \equiv \vec r_i - \vec r_j \; .
$$
The resonant medium has cylindrical shape of radius $R$ and length $L$. 
The transition wavelength $\lambda$ is such that
\be
\label{33}
\frac{\lambda}{R} \ll 1 \; , \qquad
\frac{R}{L} \ll 1 \; .
\ee
It is convenient to pass to a continuous space variable $\vec r$ by 
transforming the sums in integrals according to the rule
$$
\sum_{i=1}^N f_i = \rho \int f(\vec r)\; d\vec r \qquad
\left ( \rho \equiv \frac{N}{V}\right ) \; .
$$
Then the evolution equations for the statistical averages
\be
\label{34}
u(\vec r,t) \equiv <\sgm^-(\vec r,t)> \; , \qquad
s(\vec r,t) \equiv <\sgm^z(\vec r,t)>
\ee
satisfy partial integro--differential equations. Because of the inequalities
\be
\label{35}
\frac{\gm_2}{\om_0} \ll 1 \; , \qquad \frac{|\Delta|}{\om_0} \ll 1\; ,
\ee
where $\Delta\equiv \om -\om_0$ is the detuning parameter, the function 
$u$ is fast in time as compared to the slow function $s$.

The solutions to the evolution equations describe a bunch of filaments 
aligned along the $z$ axis which is the axis of the sample. Each filament 
has a radius $R_f$ and is centered at a point $\{ x_n,y_n\}$ in the 
transverse cross--section. The location of the filament centers is 
distributed chaotically. Thus, the solutions may be presented as 
expansions over filaments in the form
$$
u(\vec r,t) = \sum_{n=1}^{N_f} u_n(t) e^{ikz} \;\Theta\left ( R_f -
\sqrt{(x-x_n)^2 + (y-y_n)^2} \right ) \; ,
$$
\be
\label{36}
s(\vec r,t) = \sum_{n=1}^{N_f} s_n(t) \; \Theta\left ( R_f -
\sqrt{(x-x_n)^2 + (y-y_n)^2} \right ) \; ,
\ee
where $N_f$ is the number of filaments and $\Theta(\cdot)$ is a unit-step 
function. The number of filaments is related to the filament radius $R_f$ 
and the pumping characteristique
\be
\label{37}
\zeta(t) = \frac{1}{V}\int s(\vec r,t) \; d\vec r \; ,
\ee
which yields
\be
\label{38}
N_f = \frac{1}{2}\; ( 1 + \zeta)\left ( \frac{R}{R_f}\right )^2 \; .
\ee
The filament radius $R_f$ can be defined according to the procedure of 
subsection 2.5. To this end, $R_f$ is considered as a control function 
parametrizing the filamentary space structure. It is possible to 
construct the average energy (18) corresponding to the Hamiltonian (31). 
Minimizing this energy functional with respect to $R_f$ gives
\be
\label{39}
R_f = 0.22\; \sqrt{\lambda\; L} \; .
\ee
Then the number of filaments (38) can be written as
\be
\label{40}
N_f = 3.3\; (1 +\zeta )F \qquad \left ( F \equiv \frac{\pi R^2}{\lambda L}
\right ) \; ,
\ee
where $F$ is a Fresnel number.

The predictions of the theory [39--42] have been found to be in a good 
agreement with measurements, as has been confirmed in a series of 
experiments [29--34] with different lasers.

\section{Semiconfinement of Neutral Atoms}

Dynamics of neutral atoms in nonuniform magnetic fields concerns problems 
of current experimental and theoretical interest, especially with regard 
to atoms in magnetic traps, where the atoms can be cooled down to 
experience the Bose--Einstein condensation [43,44]. The motion of confined 
atoms is usually described by means of the adiabatic approximation 
assuming that the atom spins are permanently aligned along the local 
magnetic field and, thus, adiabatically follow its direction. To study 
the more general case, when atoms are permitted to escape from a trap, 
one has to invoke a more refined approximation, such as the scale 
separation approach described in Section 2.

The motion of neutral atoms in magnetic fields can be presented by the 
semiclassical equations for the quantum--mechanical average of the 
real--space coordinate, $\vec R=\{ R_\al\}$, and for the average 
$\vec S=\{ S_\al\}$ of the spin operator, with $\al=x,y,z$. The first 
equation writes
\be
\label{41}
\frac{d^2R_\al}{dt^2} = \frac{\mu_0}{m}\;\vec S\cdot
\frac{\prt\vec B}{\prt R_\al} + \gm\xi_\al \; ,
\ee
where $\mu_0$ is magnetic moment, $m$ is mass of an atom, $\vec B$ is a 
magnetic field, and $\gm\xi_\al$ is a collision term. The equation for the 
average spin is
\be
\label{42}
\frac{d\vec S}{dt} = \frac{\mu_0}{\hbar}\;\vec S\times\vec B \; .
\ee
The total magnetic field $\vec B=\vec B_1+\vec B_2$ consists of two 
terms. One is the quadrupole field
\be
\label{43}
\vec B_1 = B_1'\left ( R_x\vec e_x + R_y\vec e_y +
\lambda R_z\vec e_z\right ) \; ,
\ee
in which $\lambda$ is the anisotropy parameter. The second term is a 
transverse field
\be
\label{44}
\vec B_2 = B_2\left ( \vec e_x\cos\om t + \vec e_y\sin\om t\right ) \; .
\ee

The characteristic frequencies
\be
\label{45}
\om_1 \equiv \left ( \frac{\mu_0B_1'}{mR_0}\right )^{1/2} \; , \qquad
\om_2 \equiv \frac{\mu_0B_2}{\hbar} \; ,
\ee
where $R_0\equiv B_2/B_1'$, satisfy the inequalities
\be
\label{46}
\frac{\om_1}{\om_2} \ll 1 \; , \qquad \frac{\om}{\om_2} \ll 1 \; .
\ee
Because of the latter, the spin variable $\vec S$ has to be classified as 
fast, compared to the slow atomic variable $\vec R$.

The collision term in Eq. (41) contains a collision rate $\gm$ and a 
random collision variable $\xi_\al(t)$ defined by the stochastic averages
\be
\label{47}
\ll \xi_\al(t) \gg\; = 0 \; , \qquad \ll\xi_\al(t)\xi_\beta(t')\gg \; =
2D_\al\delta_{\al\beta}\delta(t-t') \; ,
\ee
where $D_\al$ is a diffusion rate.

The semiconfining regime of motion can be realized by preparing for the spin
variable nonadiabatic initial conditions
\be
\label{48}
S_x^0 = S_y^0 = 0 \; , \qquad S_z^0 \equiv S \neq 0 \; ,
\ee
which can be done e.g. by means of an external pulse at $t=0$. Then it is 
possible to show [45--47] that the motion of atoms becomes axially 
restricted by the value 
\begin{eqnarray}
z_mR_0 =\left\{ \begin{array}{cc}
\min_t R_z(t) & (\lambda S>0) \\
\nonumber
\max_t R_z(t) & (\lambda S<0 ) \; ,
\end{array} \right.
\end{eqnarray}
such that
\be
\label{49}
z_m^3 = z_0^3 - \frac{3\dot{z}_0^2}{2\lambda^3 S\om_1^2} \; .
\ee
Atomic collisions do not disturb the semiconfined motion provided that 
temperature $T$ is sufficiently low satisfying the condition
\be
\label{50}
\frac{k_BT\;\hbar\;\rho^2\;a_0^2}{m^2\;\om_1^3} \ll 1 \; ,
\ee
where $\rho$ is the density of particles and $a_0$ is a scattering length.

The semiconfining regime of motion makes it possible to form 
well--collimated beams of neutral particles by means of only magnetic 
fields. This regime can be employed for creating coherent beams of Bose 
atoms from atom lasers.

\section{Negative Electric Current}

Semiconductors with nonuniform distribution of charge carriers can 
exhibit rather unusual transport properties. For example, in a sample 
biased with an external constant voltage, the transient 
effect of negative electric current can happen [8,9].

Transport properties of semiconductors are usually described by the 
semiclassical drift--diffusion equations. In what follows a plane device is
considered and all quantities are expressed in dimensionless form, so 
that the space variable is $x\in [0,1]$. The continuity equation writes
\be
\label{51}
\frac{\prt\rho_i}{\prt t} + \mu_i\; \frac{\prt}{\prt x}\; ( \rho_i E) -
D_i\; \frac{\prt^2\rho_i}{\prt x^2} +\frac{\rho_i}{\tau_i} = 0 \; ,
\ee
where $\rho(x,t)$ is a charge density; $E(x,t)$ is the electric current;
$\mu_i,\; D_i$, and $\tau_i$ are mobility, diffusion coefficient, and 
relaxation time, respectively, for the carriers of type $i$. The Poisson 
equation is 
\be
\label{52}
\frac{\prt E}{\prt x} = 4\pi \sum_i \rho_i \; .
\ee
At the initial time, the distribution of charge carriers is nonuniform, 
given by
\be
\label{53}
\rho_i(x,0) = f_i(x) \; .
\ee
The sample is biased with an external constant voltage, which means that
\be
\label{54}
\int_0^1 E(x,t) \; dx = 1\; .
\ee
The total electric current through the semiconductor sample is
\be
\label{55}
J(t) \equiv \int_0^1 j(x,t) \; dx \; ,
\ee
where the density of current
$$
j = \sum_i \left ( \mu_i E - D_i\;\frac{\prt}{\prt x}\right )
\rho_i + \frac{1}{4\pi}\;\frac{\prt E}{\prt t} \; .
$$

Owing to the voltage integral (54), one has
\be
\label{56}
\int_0^1 \frac{\prt}{\prt t}\; E(x,t) \; dx = 0 \; .
\ee
It is also possible to show that
\be
\label{57}
\lim_{\tau\ra\infty} \frac{1}{\tau} \int_0^\tau \frac{\prt}{\prt x} 
E(x,t) \; dt = 0 \; .
\ee
This means that the function $E$ can be considered as slow on average in 
time and space. Then, treating $E$ as a quasi--invariant, one may find 
the solutions to Eqs. (51) and (52) in order to analyse their general 
space--time behaviour.

Negative electric current can appear only when the initial charge 
distribution is essentially nonuniform. If this initial distribution 
forms a narrow layer located at the point $x=a$, then the current (55) 
becomes negative for some short time close to $t=0$, if one of the 
following conditions holds true:
$$
a< \frac{1}{2} - \frac{1}{4\pi Q} \qquad 
\left ( Q > \frac{1}{2\pi}\right )\; ,
$$
\be 
\label{58}
a> \frac{1}{2} + \frac{1}{4\pi |Q|} \qquad 
\left ( Q < - \frac{1}{2\pi}\right )\; ,
\ee
where
\be
\label{59}
Q \equiv \sum_i Q_i \; , \qquad Q_i \equiv \int_0^1 \rho_i(x,0) \; dx \; .
\ee
The effect of the negative electric current can be employed for various 
purposes, as is discussed in Refs. [8,9]. For instance, when the initial 
charge layer is formed by an ion beam irradiating the semiconductor 
sample, then the location $a$ corresponds to the mean free path of these 
ions. In this case, by measuring the negative current $J(0)$, one can 
define this mean free path
\be
\label{60}
a = \frac{1}{2}  - \frac{1}{4\pi Q} \left [ 1 -\frac{J(0)}{\sum_i\mu_iQ_i}
\right ] \; .
\ee
This formula is valid for both positive and negative values of $Q$.

\section{Collective Liberation of Light}

One more interesting physical effect that has been described using the
scale separation approach is collective liberation of light [48]. Consider
an ensemble of resonant atoms which are doped into a medium with well
developed polariton effect [49], when in the spectrum of polariton states
there is a band gap. If an atom with a resonance frequency inside the
polariton gap is placed into such a medium, the atomic spontaneous emission
is suppressed, which is called the localization of light [50]. However, a
system of resonant atoms inside the polariton gap can radiate if their
coherent interaction is sufficiently strong. Thus the suppression of
spontaneous emission for a single atom can be overcome by a collective
of atoms radiating coherently. Conditions when such a collective liberation
of light can arise and the dynamics of this liberation are analysed in
Ref. [48].

In conclusion, a general method has been developed for treating strongly
nonequilibrium processes in statistical systems. This method, called the
scale separation approach, is especially useful for describing collective 
phenomena in the interaction of radiation with matter. To emphasize the 
generality of the approach, it is illustrated here by several different 
physical examples. The common feature of all considered systems is that 
their evolution is described by nonlinear differential or 
integro--differential equations. Such equations, as is known, are 
difficult to solve. The scale separation approach makes it possible to 
find accurate approximate solutions. The accuracy of the latter has been 
confirmed by numerical calculations and by comparison with experiment, 
when available. Using this approach several interesting physical problems 
have been solved and new effects are predicted.
\end{sloppypar}

\end{document}